\documentclass[twocolumn,10pt]{article}

\usepackage{amsmath}
\usepackage{color}
\usepackage{graphicx}
\usepackage{url}
\usepackage[mathscr]{euscript}

\newcommand\p{\mathcal{P}}

\newenvironment{proof-of-theorem}[1][Proof of Theorem]{\begin{trivlist}
\item[\hskip \labelsep {\bfseries #1}]}{\end{trivlist}}

\newcommand\cut[1]{}
\newcommand\mathcut[1]{}

\date{}

\begin{document}

\title{Network Clustering Approximation Algorithm Using One Pass Black Box Sampling}

\author{Thomas DuBois\\tdubois@cs.umd.edu\\University of Maryland \and Jennifer Golbeck \\ golbeck@cs.umd.edu\\University of Maryland \and Aravind Srinivasan\\srin@cs.umd.edu\\University of Maryland}

\maketitle

\begin{abstract}
Finding a good clustering of vertices in a network, where vertices in the 
same cluster are more tightly connected than those in different clusters, is a useful, important, and well-studied task.  Many clustering algorithms scale well, however they are not designed to operate upon internet-scale networks with billions of nodes or more.  We study one of the fastest and most memory efficient algorithms possible -- clustering based on the connected components in a random edge-induced subgraph.  When defining the cost of a clustering to be its distance from such a random clustering, we show that this surprisingly simple algorithm gives a solution that is within an expected factor of two or three of optimal with either of two natural distance functions.  In fact, this approximation guarantee works for any problem where there is a probability distribution on clusterings.  We then examine the behavior of this algorithm in the context of social network trust inference.

\end{abstract}

\section{Introduction}

Finding clusters or communities is one of the most important steps in network analysis.  Clusters should have high internal connectivity and relatively low connectivity with the rest of the network.  Finding such groups of similar or tightly connected vertices increases our understanding of the underlying graph~\cite{Leskovec2009community,Leskovec2008Statistical,Pereira-leal2004,Newman2004,Basak1988}, and many algorithms exist for clustering networks~\cite{ Bader2003,bandyopadhyay2003,Rattigan2007,Jiang2010,Li2010,Jin2009}.  Because the networks we work with grow all of the time, some of these algorithms are specifically designed to perform efficiently on large networks.  We take this goal to its extreme by proposing a randomized network clustering algorithm which queries each edge at most once.  We then derive approximation guarantees for the resulting clusterings and demonstrate its behavior on a number of real social networks.

While our algorithm applies to networks from any number of domains (the internet, biological networks, etc.), our primary motivation comes from using inferred trust in social networks.  With hundreds of millions of users on social networking websites and millions of pages of user-generated content coming on line every day, there are vast networks of users, content, and meta-data.  Access to this type of information is extremely powerful. There is potential to personalize and enhance users' experiences and improve our understanding of users and their behavior. In particular, connecting social network data - especially trust - to user-generated content allows systems to direct users to the most trustworthy users and data. This may be through recommender systems, search personalization, or direct presentation of trust information about other users. 

Clustering is an important challenge in this context. All the applications discussed above, and many more, can benefit from clustering over these networks. 
Motivated by these applications - particularly the problem of trust inference - our research addresses the issue of clustering the vertices in graphs.  

Using random graphs as a model, our goal is to find a clustering where vertices in a cluster are likely to be in the same connected component while vertices in different clusters are not.   DuBois et al.~\cite{dubois2009rigorous}  define the distance between two nodes to be the logarithm of the reciprocal of
the probability that they are connected. Because computing this probability exactly is intractable ($\#P$-complete~\cite{valiant:410}), they repeatedly sample random graphs to estimate 
such probabilities to within any desired precision and confidence.  If edges are chosen independently, this distance is a metric, and any one of a number of clustering algorithms can be applied.  They show that this technique works 
well in some practical settings; however it has some drawbacks -- 
most notably that many samples of the random graph are required to accurately estimate distances between nodes, and hence the running time involved may 
be prohibitive for very large graphs.  On the Web, where interesting graphs tend to be large, this is a major issue.

In this paper, we present a new method for graph clustering where every edge is mapped to an independent probability of its being in an instance of the graph.  The connected components of the resulting graphs, which we can sample with a depth first search, are its clusters.  
Our algorithm is computationally efficient - only a single pass 
is needed.  Furthermore, it applies not only to network clustering, but to any problem where clusterings come from any probability distribution which we can sample.

To analyze this algorithm, we define a distance function between any two clusterings and attempts to minimize the expected distance between its output and a randomly sampled clustering. We show that good clusterings can be found in expectation directly by sampling the random graph only once.  We then show that repeated sampling improves our confidence in the result.  In Section~\ref{sec:formal} we formalize the problem and prove that a single random sample gives a 3-approximation in expectation.  In Section~\ref{sec:whp} we show how to use multiple samples to improve on our probabilistic guarantees.  Finally in Section~\ref{sec:applied} we apply our new algorithm to trust inference clustering as a demonstration of its usefulness. 

\section{Related Work}

We begin our literature review with an overview of our target application -- social network trust inference and the usage of trust-based clusters, and then move on to a discussion of other clustering algorithms.

Since an individual in a social network usually knows only a tiny fraction of all the users, it is important to have some mechanism for estimating the relative importance of unknown users. In many web-based applications that seek to personalize the user's experience, this will take the form of computing their influence or trustworthiness. Trust propagation is a particularly challenging problem because of the many social and interpersonal factors that play into trust.

There are many trust inference algorithms that take advantage of given trust values and the structure of a social network, including Advogato~\cite{advogato}, Appleseed~\cite{appleseed}, Sunny~\cite{1754397}, and Moletrust~\cite{moleskiing}. These algorithms use trust that is assigned on a continuous scale (e.g. 1-10).
Trust can also be treated as a probability. This approach has been used in a number of algorithms, including~\cite{hang2008adaptive,dubois2009rigorous,patel2005probabilistic,j¿sang2006exploring}. 
The difficulty of generating these probabilities, using influence as a proxy for trust, was addressed in~\cite{1718518}. In our research, we work with probabilities that are given \emph{a priori}, but those derived from other 
methods could also be used in our algorithms. 

The result of these algorithms have a wide range of applications. Recommender systems are a common application, where computed trust values are used in place of traditional user similarity measures to compute recommendations (e.g.~\cite{o2005trust,avesani2005trust,golbeck2006generating}). In~\cite{1718504}, the authors present a technique for using trust to estimate the \textit{truth} of information that is presented, which in turn has applications for assessing information quality, particularly on the Semantic Web. More specific applications of that idea include using trust for semantic web service composition~\cite{wsc}.

Often these algorithms require, as an intermediate step, finding clusters of people who are more tightly connected to each other than to the remainder of the population~\cite{sarwar02recommender,Recommender2009}.  The art of finding useful sets of clusters has been well studied on a wide range of applications.  In some cases there is some (unknown) ``ground-truth'' clustering inherent in the data which we want to find, and the algorithms attempt to find a clustering that is ``close'' to the true one~\cite{Dasgupta99,Balcan2009}.  Often, though, there is no reason to believe that the data has inherently correct clusters, and the goal becomes simply to produce a clustering which works well in practice for a particular application.  

When each data point to be clustered consists of a vector of numerical values, one common technique is to choose a distance function between the elements (Euclidean, L1-norm, etc.) and look for clusters which minimize some optimization function.  Examples of these algorithms include k-means~\cite{kmeans79} (which minimizes the 
mean squared-distance of elements from their cluster centers), and k-centers~\cite{kcenter85} (which minimizes the maximum distance from any point to the center of a cluster).  Typically approximation algorithms, which find solutions close to optimal, are used because it is impractical to compute the optimal clustering for these problems.  For a more extensive overview of various clustering algorithms, see~\cite{ClusterSurvey2005}.   

Much work has also been done specifically on clustering networks, and we give an overview of it here.  
Newman and Girvan~\cite{Newman2004} compute the shortest path between all pairs of nodes in the network, remove the edge used in the most such paths, and repeat.  If an edge is contained in many shortest paths, then intuitively there are not many other short paths around it, and it may be a bridge between natural clusters.  Some edge removals will disconnect a component in the graph.  The order in which components disconnect gives a hierarchical clustering of the network.  They can then choose the level of hierarchy which best suites some application-specific optimization function.  Because repeatedly computing the shortest path between all pairs may not be efficient enough, Tang et al. 2011~\cite{Tang2011} build on this work by using center distance to zone~\cite{Tang2009} as a more efficient approximation of shortest paths. Several techniques pick a node at random and attempt to build up a cluster around it by repeatedly adding similar nodes.  Xu et al.~\cite{Xu2007} find structural clusters (clusters which many have low density, but have a core backbone of nodes whose neighborhoods overlap greatly with other core nodes).  Jiang and Singh~\cite{Jiang2010} propose a similar algorithm for clustering biological networks where at every step a currently active cluster expands by adding the ``closest'' node if its proximity exceeds some threshold.  

Frequently we cluster networks in order to find inherent communities in the data.  Leskovec et al. perform an extensive study on the best communities of different sizes in many large social networks~\cite{Leskovec2009community}.  They use conductance (or the normalized cut metric~\cite{Shi1997}), defined as the ratio of edges between the community and the outside world to edges within the community, as a measure of community strength.  For all of the networks they examine, regardless of size, maximum community conductance drops off considerably for community sizes greater than one hundred.  This results suggests that there may be no clusterings of large social networks which help us understand the networks structure.  However even if clustering such networks does not reveal anything important about them, it may still be useful in getting better application-specific results or efficiency.

\section{The Algorithm}

Recently DuBois et al.\cite{dubois2009rigorous} proposed an interpretation of trust within a social network based upon taking a random edge-induced subgraph.  In their framework, the direct trust on an edge corresponds to the probability that the edge will be in a random instance of the graph and indirect trust between any two people in the network corresponds to the probability that there is a path between them.  The ability to cluster the network into groups of relatively high trust ranks among their main contributions.  They find these clusters by repeatedly sampling the random graph to estimate the path probability between all pairs of nodes, and then apply various well-studied clustering algorithms to the resulting distances.  In order to have confidence that all of these pairwise distances are accurate to within a tolerance of $\pm \delta$, $O(\frac{\log n}{\delta^2})$ samples are needed on a network with $n$ people.  This poses a major drawback on internet-scale datasets, which can have millions or billions of users.  

Their solution takes a trust network, computes a distance function between pairs of points, and then uses those distances to find a clustering.  This solution scales fairly well, but we would like to do better.  There is nothing inherent in clustering which requires computing distances as an intermediate step, which inspired us to skip the distance computation altogether.  Our algorithm takes a single sample of a random graph and uses its connected component decomposition as the clustering.  Sampling the graph and computing the connected components can be done simultaneously in a single pass over the edges of the graph using depth first search, and thus is as fast of an algorithm as we can reasonably hope for.

Of course a fast algorithm that produces poor results would not be useful.  In Section~\ref{sec:formal} we derive probabilistic bounds on the quality of the resulting clusterings.  We start by defining a distance function between clusterings with the goal of minimizing the distance between our chosen clustering and a connected component decomposition of an instance of the random graph.  We do not expect to be able to find the best such clustering (and in the general case where sample clusters come not from a random graph but from a black box it is not possible), however our single sample algorithm achieves a 3-approximation in expectation.  This means that a random clustering will, on average, produce distances no more than 3 times those produced by the best clustering.  We also show that using a slightly different distance function (one used by Balcan et al.~\cite{Balcan2009}) our algorithm achieves a 2-approximation in expectation.  Finally we show how to achieve improved bounds on the deviation from this expectation by sampling multiple times.

\subsection{Definitions and Formal Analysis}
\label{sec:formal}

Consider the following problem:
\begin{itemize}
\item A clustering of a set $U$ is a set $C$ of subsets of $U$ such that $\forall x\in U, |\{s\in C: x\in s\}|=1$.  In other words, a clustering is a set of disjoint subsets whose union is the entire set $U$.  For convenience we let a clustering contain an arbitrary number of distinct empty sets.
\item Given a set $U$, and a probability distribution $\p$ on clusterings of $U$, we want to find a clustering $X$ which minimizes the expected distances between $X$ and a random clustering drawn from $\p$.
\item There are many possible distance functions between clusterings, we will concentrate on the following one: for two clusterings $X$ and $Y$, define $D_X(Y)$ to be $\min_{f:Y\rightarrow X} \sum_{s\in Y} |s \cup f(s)| - |s \cap f(s)|$.  In other words, for each set in $Y$, we match it to the set in $X$ which minimizes the size of the symmetric difference between the two.  We will later consider a similar distance function proposed by Balcan et al\cite{Balcan2009}.  They let the distance between two clusters be the minimum number of elements not in the matched clusters under any matching of the clusters.  For any two clusters, the distance using our metric is at least the distance using theirs, and at most twice the distance using theirs.
\end{itemize}

Note that while we do not restrict the function $f$, the optimal choices for $f$ are bijections (when we include an appropriate number of empty sets in the two clusterings).  This follows from the observation that for any optimal $f$, and for all $s\in Y$, $s\cap f(s)$ is at least half the size of both $s$ and $f(s)$.  Otherwise it would be better to map $s$ to the empty set.  Since no two distinct $s_1,s_2\in X$ can both share more than half of the elements of a single $y\in Y$, $f$ must be one-to-one.  By mapping extra, distinct empty sets onto the remaining sets in $X$ (if any), $f$ becomes a bijections.

For any given probability distribution on clusterings $\p$ (such as the one given by the connected component decomposition of a random graph), define random variable $Y$ to be a clustering drawn from that distribution.
Let $C$ be a clustering that minimizes the expectation of $D_C(Y)$.  Since the distribution $\p$ is a black box in general, we cannot hope to find the actual clustering $C$ even if we assume the $P=NP$ (because we can not distinguish between two distributions with complete certainty).  However the following simple algorithm surprisingly gives a 3-approximation in expectation:

\emph{Take a random sample $C'$ from $\p$, and use that as the approximation.}

The analysis proceeds as follows:

Let $U,\p,$ and $C$ be given.  Define $g_X(u)$ to be the set $s$ in the clustering $X$ that contains $u$ and $f_X^Y$ to be the best function mapping clusters in $Y$ to clusters in $X$.  The expected cost of the optimal solution is $E[D_C(Y)]$
\begin{align}
\nonumber &=E[D_C(C')]\\
\nonumber &=\sum_Y Pr[Y]\cdot\sum_{s\in Y} (|s\cup f_C^Y(s)|-|s\cap f_C^Y(s)|)\\
\nonumber &=\sum_{u\in U}\sum_Y Pr[Y]\cdot \left\{\begin{array}{rl}0&f_C^Y(g_Y(u))=g_C(u)\\1 &f_C^Y(\{\})=g_C(u)\\2 &\mbox{ow.}\end{array} \right.\\
\label{eq:lowerbound}&=\sum_{u\in U} \left(\begin{array}{c}\sum_{Y:f_C^Y(g_Y(u))\neq g_C(u) \wedge g_C(u)\neq f_C^Y(\{\})} Pr[Y]\\+\sum_{Y : f_C^Y(g_Y(u))\neq g_C(u)}Pr[Y]\end{array}\right)
\end{align}

Each element $u$ adds to the total cost only if its set in $Y$ does not map to its set in $C$.  In that case it costs 1 because of the mapping from $g_Y(u)$ to $f_C^Y(g_Y(u))$, and it costs another 1 if some non-empty set in $Y$ maps to $g_C(u)$.  

The expected cost, $E[D_{C'}(Y)]$, of our approximated solution is derived in Figure~\ref{fig:derivations}.
\begin{figure*}
\begin{align}
\nonumber E[D_{C'}(Y)]&=\sum_{C'}Pr[C']\cdot\sum_Y Pr[Y]\cdot\sum_{u\in U} \left\{\begin{array}{rl}0&f_{C'}^Y(g_Y(u))=g_{C'}(u)\\1 &f_{C'}^Y(\{\})=g_{C'}(u)\\2 &\mbox{ow}\end{array} \right.\\
&\leq \sum_{C', Y} Pr[C'\wedge Y]\cdot\sum_{u\in U} \left\{\begin{array}{rl}0&f_{C}^Y(g_Y(u))=g_C(u)=f_C^{C'}(g_{C'}(u))\\1 &f_{C}^Y(\{\})=?f_C^{C'}(g_{C'}(u))\\2 &\mbox{ow}\end{array} \right. \label{eq:mappingthroughC}\\
\nonumber &= \sum_{u\in U} \left(\begin{array}{c}\sum_{C',Y: f_C^Y(g_Y(u))\neq g_C(u) \vee f_C^{C'}(g_{C'}(u))\neq g_C(u)} Pr[C'\wedge Y] +\\  \sum_{C',Y: (f_C^Y(g_Y(u))\neq g_C(u) \vee f_C^{C'}(g_{C'}(u))\neq g_C(u)) \wedge f_C^Y(\{\}) \neq f_C^{C'}(g_{C'}(u))} Pr[C'\wedge Y]  \end{array}\right)\\
\nonumber &\leq \sum_{u\in U} \left(\begin{array}{c}\sum_{C',Y: f_C^Y(g_Y(u))\neq g_C(u) \vee f_C^{C'}(g_{C'}(u))\neq g_C(u)} Pr[C'\wedge Y] +\\
\sum_{C',Y: f_C^Y(g_Y(u))\neq g_C(u) \wedge f_C^Y(\{\}) \neq f_C^{C'}(g_{C'}(u))} Pr[C'\wedge Y]+\\
\sum_{C',Y: f_C^{C'}(g_{C'}(u))\neq g_C(u) \wedge f_C^Y(\{\}) \neq f_C^{C'}(g_{C'}(u))} Pr[C'\wedge Y]
\end{array}\right)\\
\label{eq:upperbound}&\leq \sum_{u\in U} \left(\sum_{Y: f_C^Y(g_Y(u))\neq g_C(u)} 3Pr[Y] +
\sum_{Y: f_C^Y(g_Y(u))\neq g_C(u) \wedge f_C^Y(\{\}) \neq g_{C'}(u)} Pr[Y]\right)
\end{align}
\caption{Derivations showing that our algorithm gives a 3-approximation in expectation.}
\label{fig:derivations}
\end{figure*}
Where Equation~\ref{eq:mappingthroughC} maps $s\in Y$ to $s'\in C'$ if and only if they both map to the same subset in $C$.  This mapping must cost at least as much as the optimal mapping $f_{C'}^Y$.  
Dividing Equation~\ref{eq:upperbound} by Equation~\ref{eq:lowerbound} gives $E[D_{C'}(Y)/D_C(Y)] \leq 3$.

We demonstrate that this upper bound is tight with the following distribution on clusterings:
$$Pr[Y=\{\{1\},\{2\}\}]=\frac{k-1}{k}, Pr[Y=\{\{1,2\}\}]=\frac{1}{k}.$$
The optimal solution simply matches the high probability case, $C=\{\{1\},\{2\}\}$.  The expected cost of this solution is $((k-1)\cdot 0 + 1\cdot 1)/k$.  The expected cost of using a random sample is $$\frac{(k-1)\cdot \left( \frac{k-1}{k}\cdot 0 + \frac{1}{k}\cdot 1\right) +1\cdot\left(\frac{k-1}{k}\cdot 2 + \frac{1}{k}\cdot 0\right)}{k}$$ which reduces to $3 \cdot \frac{k-1}{k^2}$, and thus approaches 3 times optimal as $k\rightarrow \infty$.

\subsection{Expected 2-Approximation}

We briefly consider the case where we change the distance function between clusterings to $$D_C(Y) = \min_f |\{u : f(g_Y(u))\neq g_C(u)\}|$$ (as used by~\cite{Balcan2009}) and keep all definitions and notations the same as above.  This distance function costs exactly 1 for each element $u$ whose set in $Y$ is not mapped to its set in $C$.  Using this metric, our algorithm yields a 2-approximation in expectation.  We show this by rewriting the distance function as $$E[D_C(Y)] = \sum_u \sum_{Y:f_C^Y(g_Y(u))=g_C(u)} \Pr[Y].$$  Balcan et al.~\cite{Balcan2009} observe that this function is symmetric and obeys the triangle inequality.  Therefore the expected distance $E[D_{C'}(Y)] \leq E[D_C(C')+D_C(Y)] = 2 E[D_C(Y)]$.

\subsection{Multiple Samples}
\label{sec:whp}

Depending on the application, the guarantee of a 3-approximation in expectation may not be sufficient.  An unlikely sample could have arbitrarily bad behavior.  For example, the probability that a sample is better than a 5-approximation is not guaranteed to be any higher than $1/2$.  In this subsection we explore various ways to use multiple samples to achieve better approximation guarantees.

Since our approximation guarantee is in expectation, it is important to limit the probability that we choose a bad clustering $C'$ (one where $E[D_{C'}(Y)]$ is much greater than $3E[D_C(Y)]$.  We do this using Markov's inequality.  Since the approximation ratio is always at least 1 (no solution can be better than the optimal solution),  \begin{align*}Pr[E[D_{C'}(Y)] >(3+2\epsilon)E[D_C(Y)] \leq 1/(1+\epsilon).\end{align*}  We could attempt to bound the variance in the approximation ratio, however as the above example illustrates (or any example where most of the probability mass lies close to the optimal solution, with a small amount of mass at a large distance), it can be quite bad.  
Through repeated sampling, we can do much better.  Instead of taking only a single sample clustering, let us take samples $C'_1,\ldots,C'_m$ from the distribution.  The first quantity of interest is the approximation ratio $R$ achieved by the best of these samples - $\min_i D_{C'_i}(X)$.    Since the sampled $C'_i$'s are independent, \begin{align*}R&=Pr[\min_i D_{C'_i}(X)/E[D_C(X)]>3+2\epsilon]\\& \leq \prod_i Pr[D_{C'_i}(X)/E[D_C(Y)]>3+2\epsilon]\\& \leq 1/(1+\epsilon)^m.\end{align*}  
Thus if we want at most a $\tau$ probability of having no samples within this distance, we need $m=\lceil \log_{1+\epsilon} 1/\tau\rceil$ or more samples.

The existence of a sample $C'_i$ which is close to a 3-approximation does not directly imply that we can determine which of the sample(s) are good.  If our application allows us to test each of the samples and choose one with the best results, we may not need to find the one with the best approximation ratio.  Otherwise, to be certain which of the $C'_i$ gives the best approximation ratio, we would need to know $C$ already (or at least be able to calculate $D_C(X)$ knowing $X$ but not $C$).  We can get around this by taking $l$ additional samples $\{X_1,\ldots,X_l\}$, and computing for each of the $C'_i$, the total distance from the $X_j$'s.  We then select the $C'_i$ with the minimum such total distance.  

We must now consider that the samples $X_1,\ldots,X_l$ that we draw might give a total distance larger than its expectation for the ``good'' $C'_i$ and smaller than the expectation for a ``bad'' $C'_i$.  To address this, we now show that when $l$ is sufficiently large, with high probability even if we don't select the best $C'_i$, we will select one which is close enough.  For each $C'_i$, the expected cost $E[D_{C'_i}(X)] \geq E[D_C(X)]$, and with probability at least $1-\tau$, for at least one such $C'_i$, $E[D_{C'_i}(X)]\leq (3+2\epsilon)\cdot E[D_C(X)]$.   We define $d_i=\sum_{j=1}^l D_{C'_i}(X_j)/|U|$.  Since the $d_i$ are a sum of independent random variables taking on values in $[0,1]$, we can apply Chernoff bounds to their deviation probability.  If we take $l=(|U|/\delta ^2)\cdot O(\log kp)$, then with probability at least $1-p$ all candidate's distance totals $d_i$ will be estimated to within $(1\pm\delta)$.  

If there exists a candidate with distance total $$D_{C'_i}(Y)\leq E[D_C(Y)]\cdot (3+2\epsilon) \cdot l,$$ and all such estimates are within $(1\pm\delta)$ of the true totals, then the candidate with the minimum estimated total has a true total at most $\cdot(3+2\epsilon)E[D_C(Y)]\cdot l\cdot(1+2\delta)$.  This candidate gives an approximation ratio of $(3+2\epsilon)\cdot(1+2\delta)$.  Such a candidate is found with probability at least $1-p-\tau$.

\section{Trust Inference Application}
\label{sec:applied}

\begin{figure}
\begin{tabular}{c}
\includegraphics[width=7cm]{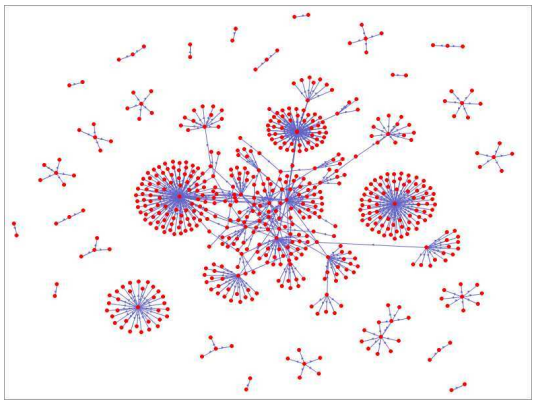}\\
{}\\
\includegraphics[width=7cm]{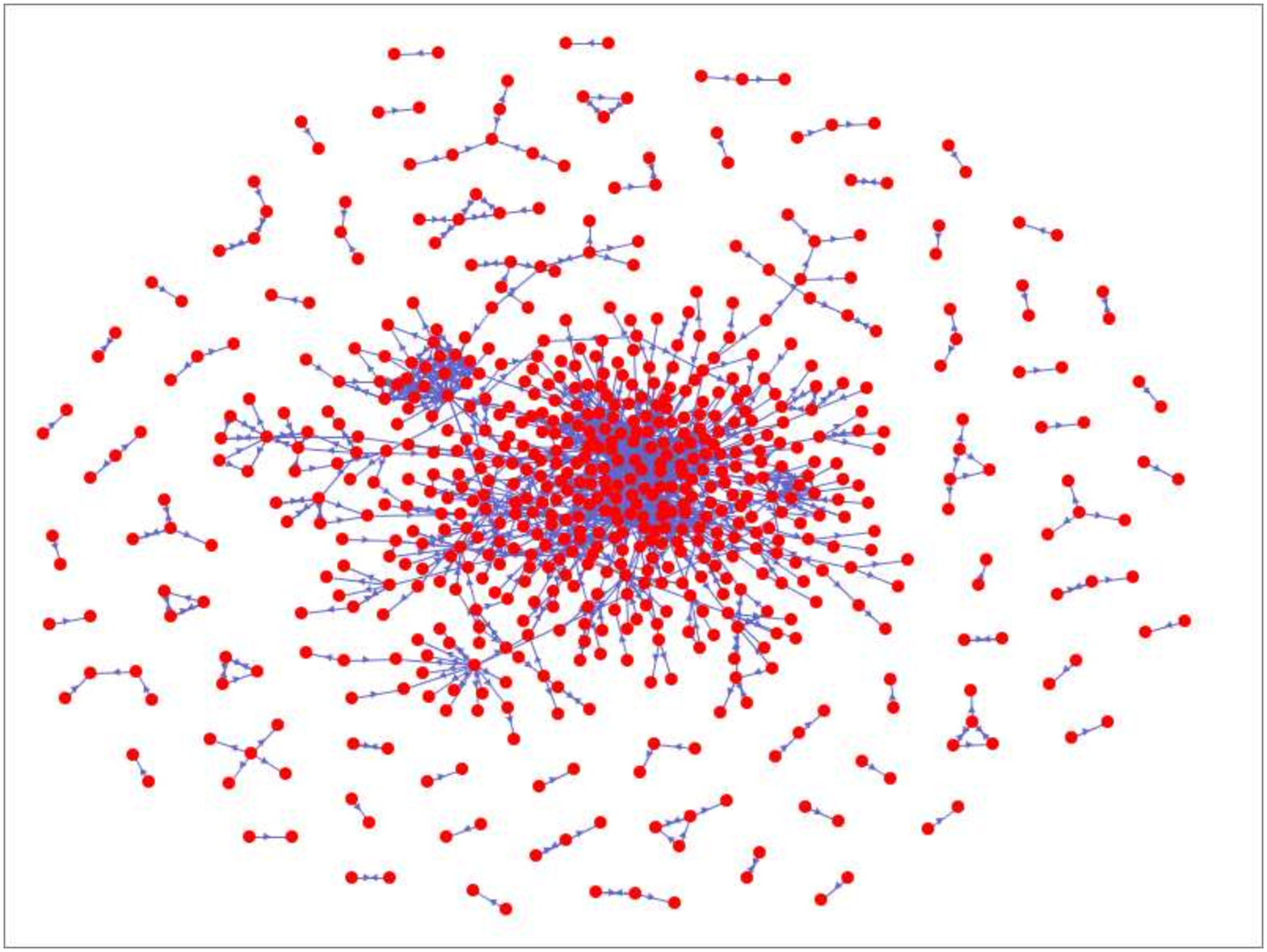}\\
{}\\
\includegraphics[width=7cm]{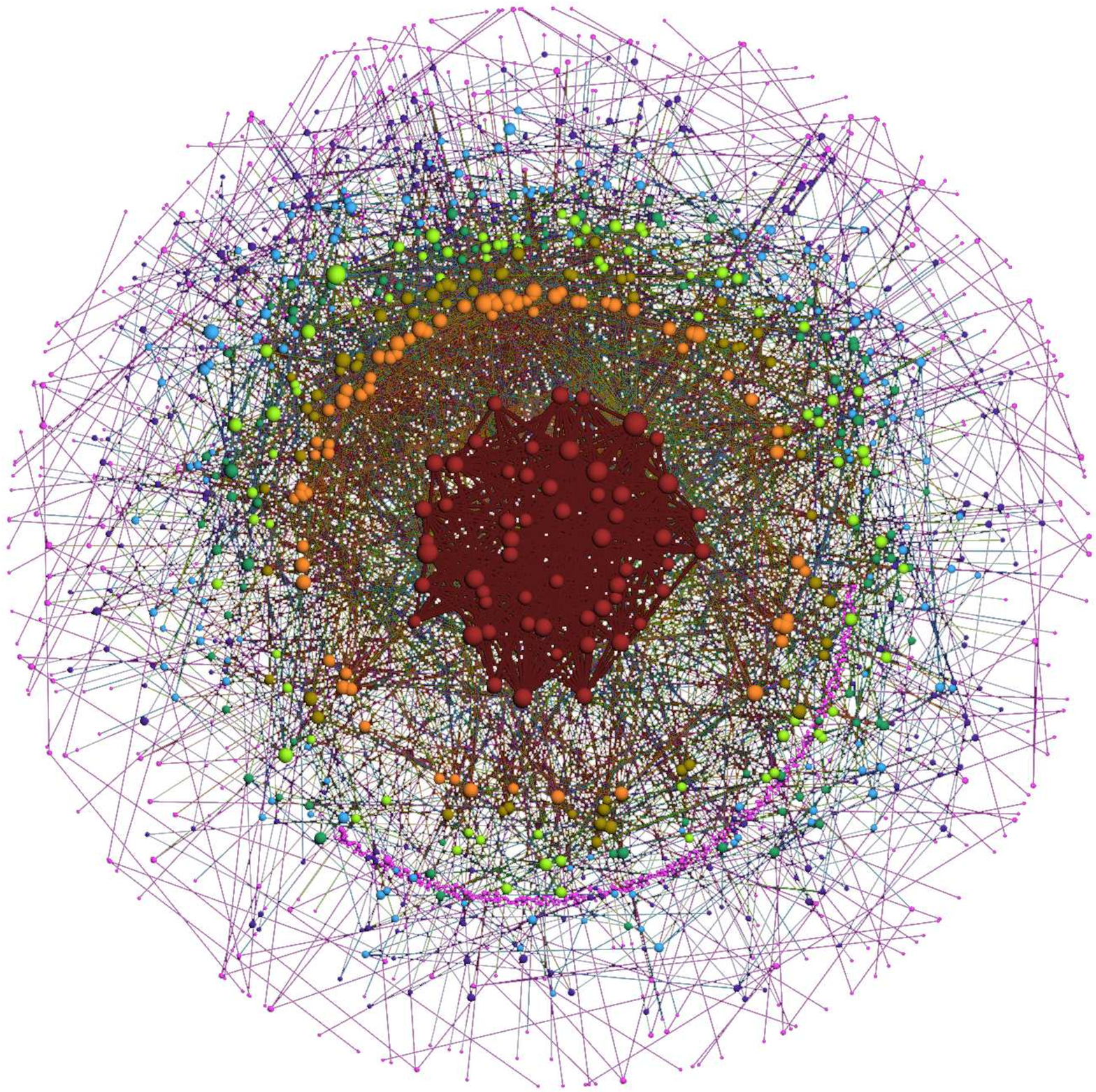}
\end{tabular}
\caption{The three networks used in our analysis have very different structures. The Trust Project Network (top) has many star formations which affect the quality of its clusters. The FilmTrust Network (middle) is a more traditionally organized social network. The Epinions Network (bottom) is much larger.}
\label{fig:nets}
\end{figure}

Having a 3-approximation algorithm is a nice theoretical result, but it does not necessarily imply practical benefits.   For example, if an optimal solution has a large expected distance (1/3 of the maximum distance for example), then a 3-approximation is meaningless.  The hope is that networks will only have such bad behavior if they are inherently not well clusterable.  There is some intuitive reason to believe this is so.  If a certain set of nodes often forms the majority of a connected component and they are in the same component $c$ of an optimal clustering, then a random clustering $Y$ will likely have a component $y$ that matches with low cost to $c$.  Meanwhile if the optimal clustering has high cost, that means that few large groups of nodes consistently form the bulk of a component.  

In this section we explore what kind of clusterings occur in real trust networks using various parameters.  We examine the Trust Project, FilmTrust\cite{FilmTrust2006}, and Epinions networks.  Visualizations of these networks are shown in Figure~\ref{fig:nets} with their sizes shown in Table~\ref{tab:nets}.  In the first two of these networks, users rate their level of trust (with respect to a specified domain) in their friends.  In the Epinions network, users rate whether they like or dislike statements made by others, and these ratings can be used as a proxy for ratings of the statements' authors. In this paper we address only positive trust, so unfavorable ratings are discarded. 

The Trust Project network is derived from an early Semantic Web trust network~\cite{diss}.  As is shown in Figure~\ref{fig:nets}, it has many star patterns. This occurs when users make connections to many friends who do not, in turn, participate in the network. Thus, they have no outgoing connections. This affects our ability to cluster the network. The FilmTrust network is built from a social network in which users rate movies and how much they trust their friends in that context. As the visualization shows, it has a more traditional network structure. However, there are a number of small groups that are disconnected from the giant component. These are shown as the small subnetworks, often pairs, floating around the edge of the visualization. Finally, the Epinions network shows social network connections on the product review site. Trust ratings indicate how much they trust one another's reviews. 

\begin{table}
\begin{center}
\begin{tabular}{|l|r|r|r|}
\hline
\textbf{Network}&\textbf{Nodes}&\textbf{Edges}&\textbf{Density }\\
\hline
Trust Project &62&105&0.055\\
\hline
FilmTrust&310&774&0.016\\
\hline
Epinions&114,467&717,667&0.0001\\
\hline

\end{tabular}
\end{center}
\caption{The size and descriptive statistics of our three example networks.  Density is calculated as the ratio of edges to possible edges.}
\label{tab:nets}
\end{table}

In all of our networks, ratings form directed trust edges.  Our first step is to create an undirected and normalized trust graph.  We convert every lone directed edge into an undirected edge, and whenever two people rate each other, we average their ratings to form a single undirected edge.  The edge weights are then normalized to fall between 1 and 10.  Since we need probabilities on the edges instead of weights, we introduce a global parameter $t$.  An edge with weight $w$ gets probability $\max(1,w/t)$.  Therefore when $t$ is small, edge probabilities tend to be high and connected components will be large, and for large $t$ edge probabilities and connected components will generally be smaller.

\begin{figure*}
\begin{center}
\begin{tabular}{cc}
\includegraphics[width=7.5cm]{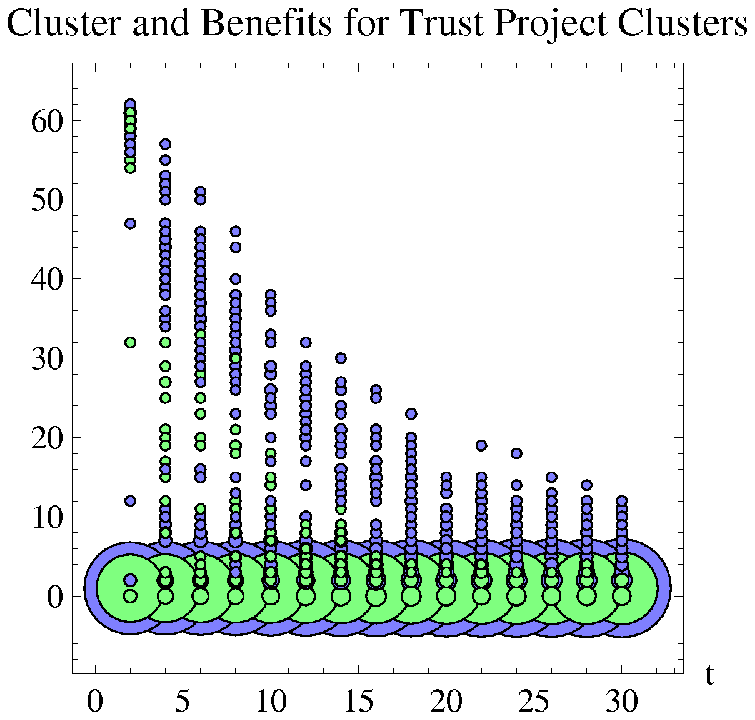}&
\includegraphics[width=7.5cm]{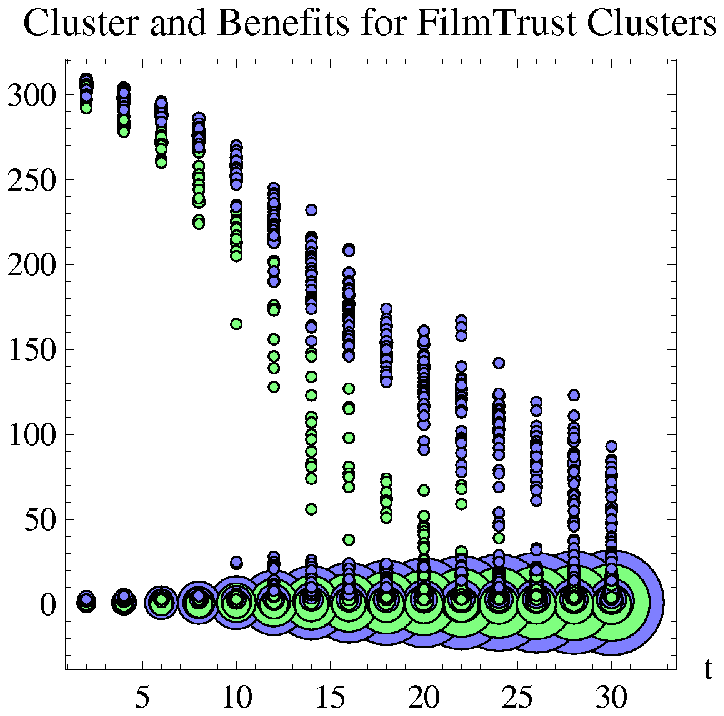}\\
\includegraphics[height=7.5cm]{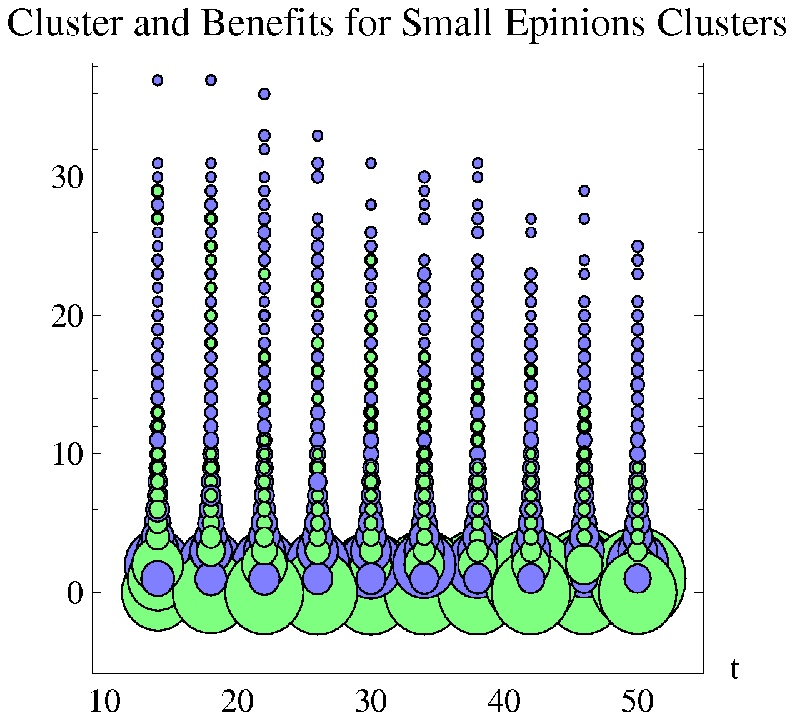}&
\includegraphics[height=7.5cm]{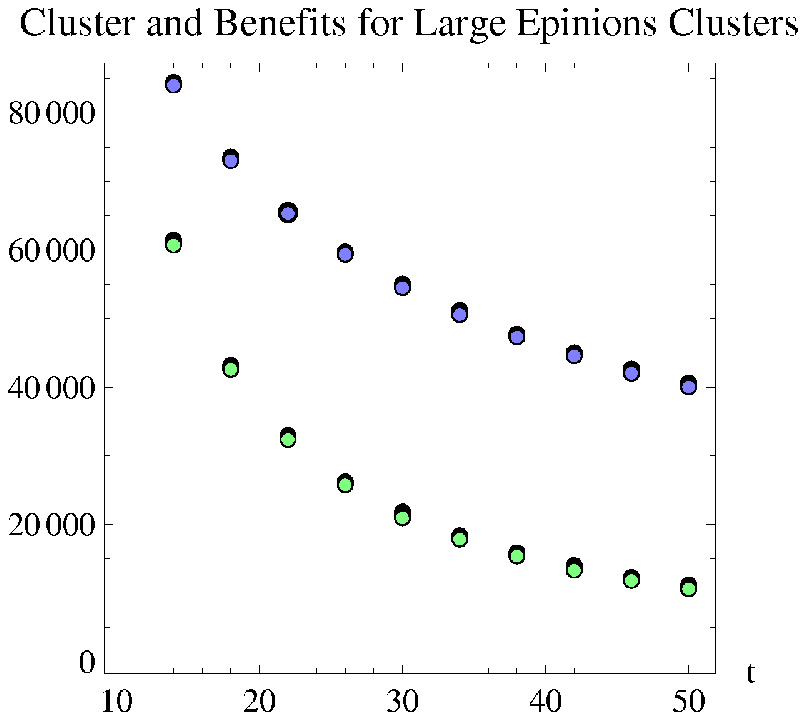}
\end{tabular}
\end{center}
\caption{The top two plots show component sizes (blue) and benefit sizes (green) within Trust Project (left) and FilmTrust (right) for $t=2$ to $30$ with $30$ samples of each.  The bottom two plots show the component sizes and benefits for the Epinions dataset, with the left plot showing small clusters and the right plot the largest clusters.
A circle centered at $(x,y)$ with radius $r$ indicates that the number of clusters of size (or benefit) $y$ with $t=x$ is proportional to $r$.}
\label{fig:compsizeft}
\end{figure*}

For the Trust Project and FilmTrust networks we vary $t$ from 2 to 30, generating 30 sample clusterings for each value.  For the Epinions network, we need considerably higher values of $t$ to capture the same behavior, so we use a range of $14$ to $50$.  In Figure~\ref{fig:compsizeft} we show the frequencies of each component size and each component's benefit, where we define the benefit of a cluster to be its size minus its cost (or how much less it costs than its maximum possible cost).  Due to our choice of cost function, two clusters each have to share at least half of their nodes to have any benefit at all.
 The x-axis shows the value of $t$, the y-axis shows the component size (or benefit), and the circle diameters show the how many components are that size (or benefit) in our samples.  This gives a sense of what size clusters to expect for different values of $t$.  Values of $t<10$ are included for informational purposes, but may be poor choices in practice, because they give equal weight to all user trust ratings $\geq t$.

\begin{figure}
\begin{tabular}{c}
\includegraphics[width=7.5cm]{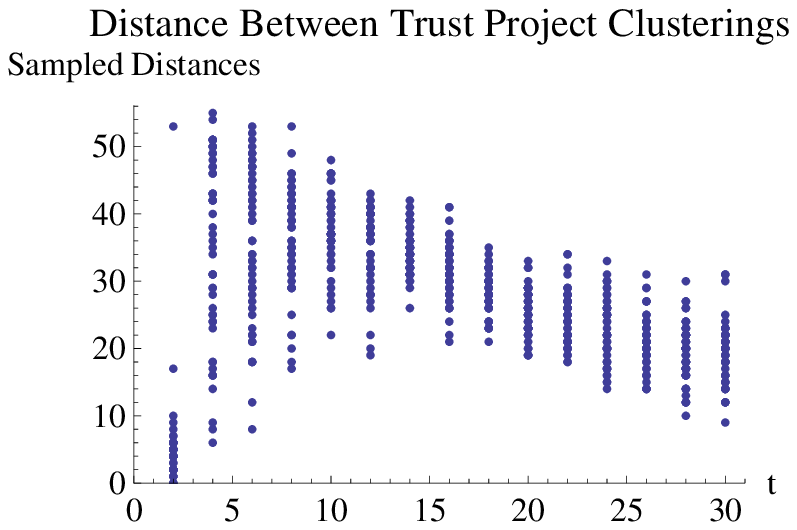}\\
{}\\
\includegraphics[width=7.5cm]{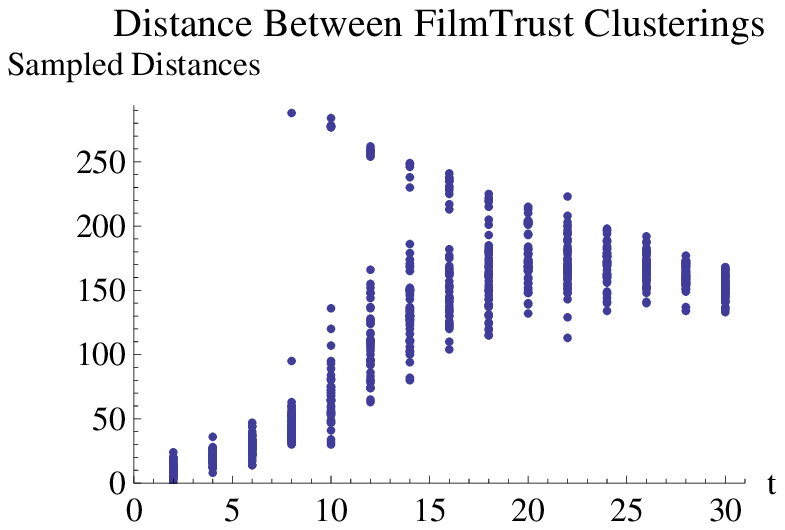}\\
{}\\
\includegraphics[width=7.5cm]{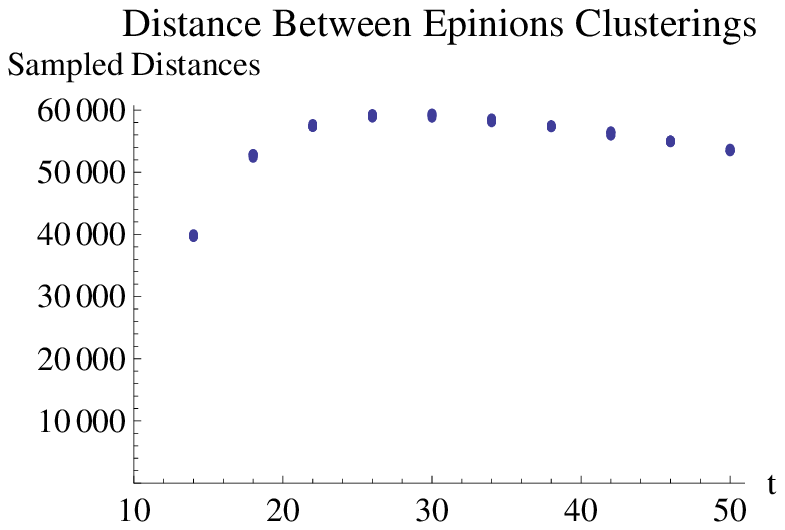}
\end{tabular}
\caption{This figure shows costs between randomly sampled clusterings for Trust Project (top), FilmTrust (middle), and Epinions (bottom) networks.  The maximum distance between samples in the smaller two networks is approximately the size of the network, whereas the maximum distance in the Epinions network is roughly half of its network size.}
\label{fig:costs}
\end{figure}

Figure~\ref{fig:costs} contains scatter plots of the distances between pairs of randomly sampled clusterings for all 3 datasets.  As discussed in Section~\ref{sec:formal}, the expected distance between two randomly chosen clusterings is at most $3$ times optimal.  So these plots demonstrate roughly how similar clusterings are, and what range $E[D_C(X)]$ falls into.
When $t\rightarrow 0$, the random graphs lose their randomness and are always connected.  Conversely at $t\rightarrow \infty$, the graphs are always disconnected.  Therefore at the two extremes distances will be $0$.  Of interest here is the shape of the curve in between, and specifically for what values of $t$ are there good, representative clusterings.

From Figures~\ref{fig:compsizeft} and \ref{fig:costs}, it is evident that the Trust Project network does not produce particularly stable clusters.  Most of the benefits, even for relatively small values of $t$, are quite small when compared to the larger cluster sizes.  This means that there is not much more than $1/2$ overlap between matched clusters.  Instead most of it's benefit comes from small clusters matching up well.  This may be (in part) a product of star shaped connections.  Under the right conditions a star graph should form a single cluster.  But with our algorithm it will form a random large cluster and many singletons, which will have high distance from another such random clustering.  

The FilmTrust network creates considerably more consistent clusterings.  Much more of the benefit comes from a large, fairly consistent cluster, but considerable benefit comes from smaller clusters as well.  Even with $t$ as high as 20 (which corresponds to maximum trust giving only a $1/2$ edge probability), there are still large clusters that share a considerable core component, indicating a very stable cluster.  For this network as well as Trust Project, the shape of the cost curves largely depends on the giant component.  If it exists and is stable for a given $t$, the costs are low.  If it exists but changes significantly with different samples, then some pairs have low cost, and some have high cost.

In this Epinions dataset, our algorithm consistently identifies a stable giant component and a number of smaller components of widely varying stability. This clustering behavior is particularly useful for applications that use the trust values to boost performance. Trust values can be used in recommender systems to generate predictive ratings based on a user's social connections~\cite{golbeck2006generating}. 
Integrating information about trust clusters into traditional recommender systems can significantly improve the accuracy of recommendations~\cite{Recommender2009}.   The smaller groups of size $<$30 that identified may reflect the types of niche interest groups that benefit most from the trust clustering recommendation techniques.

\section{Conclusion}

In this paper, we introduce a simple and extremely efficient network clustering algorithm, mathematically derive bounds on its expected approximation ratio and probability of substantial deviation from this ratio,  
 and demonstrate its application in social trust networks. We treat trust as a probability on edges in the network and present an algorithm that only 
requires the ability to sample clusters from a black-box probability distribution rather than explicitly computing distance in the network. We then prove that this  is a $3$-approximation algorithm with good theoretical performance. To show the practical applications, we test the algorithm on three real-world social trust networks. 

This work has applications for mining social networks, recommender systems, content filtering, and more. Future work will explore on what types of networks this algorithm is most effective as well as which applications can benefit the most from its.

\section{Acknowledgements}

This work was supported by NSF Award CNS 1010789, NSF Award CNS-0626636,
and U.S. Army Research Office grant W911NF1010350.

\bibliographystyle{plain}
\bibliography{clustering}

\end{document}